\documentclass[epj]{svjour}
\usepackage{color}
\usepackage[normalem]{ulem}  
\usepackage[dvipsnames]{xcolor}

\usepackage{graphics}
\begin{document}
\title{Moment of inertia, quadrupole moment, Love number of neutron star and 
their relations with strange matter equations of state}
\author{Debades Bandyopadhyay\inst{1}, Sajad A. Bhat\inst{1}, Prasanta Char\inst{2} \and Debarati Chatterjee \inst{3}
}                     
%
\mail{debades.bandyopadhyay@saha.ac.in}
\institute{Saha Institute of Nuclear Physics, HBNI, 1/AF Bidhannagar, 
Kolkata-700064, India \and Inter-University Centre for Astronomy and 
Astrophysics, Post Bag 4, Ganeshkhind, Pune - 411007 \and LPC/ENSICAEN, 6 
Boulevard Marechal Juin, 14050, Caen, France}
\date{Received: date / Revised version: date}
%
\abstract{
We investigate the impact of strange matter equations of state involving
$\Lambda$ hyperons, Bose-Einstein condensate of $K^-$ mesons and first 
order hadron-quark phase transition on moment of inertia,
quadrupole moment and tidal deformability parameter of slowly rotating neutron 
stars. All these equations of state are compatible with the 2 M$_{solar}$ 
constraint. The main findings of this investigation are the universality of the
I-Q and I-Love number relations, which are preserved by the EoSs including 
$\Lambda$ hyperons and antikaon condensates, but broken in presence of a 
first order hadron-quark phase transition. Furthermore, it is also noted
that the quadrupole moment approaches the Kerr value of a black hole for 
maximum mass neutron stars. 
\PACS{97.60.Jd neutron stars - 26.60.Kp equations of state
     } 
} 
\titlerunning{Moment of inertia, quadrupole moment ...}
\authorrunning{D. Bandyopadhyay et al.}
\maketitle
\section{Introduction}
\label{intro}
Dense matter equation of state (EoS) is a key ingredient to understanding 
physics of core-collapse supernovae, neutron stars and merger of neutron stars.
The determination of EoS of dense matter is a challenging task for the 
scientific community. Low density matter below the saturation density might be
constrained using inputs from nuclear physics experiments in laboratories. On
the other hand, the EoS at several times normal nuclear matter density is 
probed by astrophysical observations.   

Neutron stars are investigated in multi-wavelength observations. These resulted
in macroscopic properties of neutron stars, for example, masses, surface 
magnetic fields and temperatures. Mass measurement in neutron star binaries 
particularly relativistic binaries involving pulsars has reached a high 
precision level. The highest neutron star mass measured most accurately so far 
is 2.01$\pm 0.04$ M$_{solar}$ 
\cite{anto}. This puts severe constraint on the EoS of dense matter. Radius 
measurement is still a debatable issue. Efforts to extract information about
radius due to surface emission in x-ray thermonuclear bursts and from accreting 
neutron stars in quiescence are going on, but complicated by uncertainties in 
the composition of atmosphere and distance to sources. Recently launched Neutron
Star Composition Explorer (NICER) Mission could provide a better estimation of 
neutron radius observing and analysing soft x-rays from rotation powered 
pulsars if the mass is known accurately. An alternative to this is 
the measurement of moment of inertia (I), for example that of pulsar A in 
double pulsar system PSR J0737-3039 \cite{burg,schu}. There will be many fold
increase in the number of newly discovered pulsars in relativistic binaries with
the advent of highly sensitive Square Kilometre Array (SKA). This will 
facilitate faster estimation of moment of inertia in the SKA era. Consequently,
simultaneous knowledge of mass and radius of same neutron star will be 
available. Furthermore, this might lead to the better understanding of EoS in
neutron star interior.     

After the discoveries of black hole and neutron star
merger events, gravitational waves 
open up a new vista into astrophysical observations. It is expected that 
gravitational wave detectors such as Advanced LIGO, VIRGO or LIGO-India 
\cite{Indigo} might record 
gravitational waves from neutron star merger events. Gravitational waves from 
merging neutron stars provide an interesting opportunity to probe dense matter
in neutron star interior and its EoS \cite{abbott,ott17}. In the late 
inspiralling phase of binary neutron stars, tidal deformations could
be estimated. This tidal deformation depends on the 
EoS. The tidal deformation
is described by a set of parameters known as Love numbers. The quadrupole Love 
number ($k_2$) is an important quantity which is given by the ratio of
quadrupole and tidal tensors. It might be possible to extract the Love number
from the detection of gravitational wave signal \cite{flan,hind,read}. The 
recent investigation on this issue found that the Love number was related to 
moment of inertia and spin induced quadrupole moment ($Q$) through universal 
relations \cite{yagi}.

The core motivation to find a universal relation was to describe the 
exterior spacetime of a compact object independently of its internal 
structure, as the lack of experimental and observational findings results in 
poor knowledge of the interior of those objects. However, to use these 
relations one must be careful to incorporate the proper physical conditions 
where these are valid. For example, it is found that at high rotational 
frequencies $\sim 1 kHz$ \cite{doneva14} and high magnetic field 
$\sim 10^{13} G$  \cite{haskell14} the universal relations are broken. 
The usage and implications of the so called
I-Love-Q relations can be found in the recent review by Yagi and Yunes 
\cite{yagi17} and references therein. However, we find the lack of evidences
in the literature regarding whether these relations will also hold strong in 
presence of a first order phase transition inside the neutron star or not. 
This motivates us to look into this problem.

Observed masses, radii and moments of inertia are direct probes of compositions
and EoS of dense matter in neutron stars. The appearance of strange matter in
the form of novel phases of hyperons, Bose-Einstein condensate of antikaons and 
quarks are highly plausible above 2-3 times normal nuclear matter density in
neutron star interior. Strange matter whether it is hyperon matter 
\cite{gle82,sch96,weis,dc}, kaon condensed matter \cite{kap,knor} or quark 
matter \cite{farhi}, makes the EoS softer and results in reduction in 
maximum mass compared with that of neutron stars made of neutrons and protons 
only. Keeping pace with latest observations, EoS models have to
explain two solar mass or more massive neutron stars. Strange matter EoS models
encounter a precarious situation. For hyperons, this is known as the hyperon 
puzzle. This puzzle could be solved by introducing an extra repulsion into the 
EoS \cite{weis}.  

Motivated by the recent advances in multi-wavelength observations of neutron 
stars and the detection of gravitational waves in neutron star
merger, we investigate dense matter EoSs involving strangeness degrees of 
freedom and their impact on the moment of inertia, quadrupole moment and love 
number. Furthermore, we want to explore whether universal relations among those
observables hold good even in presence of exotic matter involving 
first order phase transition or not. 

We use geometrized units $c=G=1$ in this paper. 
The paper is organised in the following way. We describe dense matter EoSs with
various compositions within the framework of a field theoretical model and the
calculation of love number, moment of inertia and quadrupole moment in Section
2. Results are discussed in Section 3. We conclude in Section 4. 

\section{Equation of State}
We compute the moment of inertia, quadrupole moment and love number in slowly
rotating neutron stars. Each of these quantities is dependent on the EoS. The
EoS for charge-neutral and beta-equilibrated matter made of baryons, antikaon 
condensate and leptons is constructed.  
We adopt a density dependent relativistic hadron (DDRH) field theory for the 
description of strongly interacting dense baryonic matter. Baryon-baryon 
interaction in the DDRH model is mediated by exchanges of scalar $\sigma$, 
vector $\omega$, $\phi$ and $\rho$ mesons and is given by 
the Lagrangian density \cite{banik14,char14,typ10}, 

\begin{eqnarray}
\label{eq_lag_b}
{\cal L}_B &=& \sum_B \bar\psi_{B}\left(i\gamma_\mu{\partial^\mu} - m_B
+ g_{\sigma B} \sigma - g_{\omega B} \gamma_\mu \omega^\mu 
- g_{\phi B} \gamma_\mu \phi^\mu \right. \nonumber\\
&& \left. - g_{\rho B} 
\gamma_\mu{\mbox{\boldmath $\tau$}}_B \cdot 
{\mbox{\boldmath $\rho$}}^\mu  \right)\psi_B\nonumber\\
&& + \frac{1}{2}\left( \partial_\mu \sigma\partial^\mu \sigma
- m_\sigma^2 \sigma^2\right)
-\frac{1}{4} \omega_{\mu\nu}\omega^{\mu\nu}\nonumber\\
&&+\frac{1}{2}m_\omega^2 \omega_\mu \omega^\mu
-\frac{1}{4} \phi_{\mu\nu}\phi^{\mu\nu}
+\frac{1}{2}m_\phi^2 \phi_\mu \phi^\mu \nonumber\\
&&- \frac{1}{4}{\mbox {\boldmath $\rho$}}_{\mu\nu} \cdot
{\mbox {\boldmath $\rho$}}^{\mu\nu}
+ \frac{1}{2}m_\rho^2 {\mbox {\boldmath $\rho$}}_\mu \cdot
{\mbox {\boldmath $\rho$}}^\mu.
\label{had}
\end{eqnarray}
Here $\psi_B$ stands for the baryon octet, ${\mbox{\boldmath 
$\tau_{B}$}}$ is the isospin operator and density dependent 
meson-baryon couplings are denoted by $g$s. The exchange of $\phi$ mesons 
accounts for the repulsive hyperon-hyperon interaction and could be a plausible
solution to the hyperon puzzle. Nucleon-nucleon interaction is not 
mediated by $\phi$ mesons.

Meson field equations are solved in the mean field approximation (MFA). Finally
we obtain the pressure versus energy density known as the EoS as given by 
Ref.\cite{banik14,char14}. For density dependent couplings, the pressure 
includes the rearrangement term which takes care of many body effects and 
thermodynamic consistency. 
\begin{center}
\begin{figure}
\vspace*{1cm}       
\hspace{0.5cm}
\resizebox{0.4\textwidth}{!}{\includegraphics{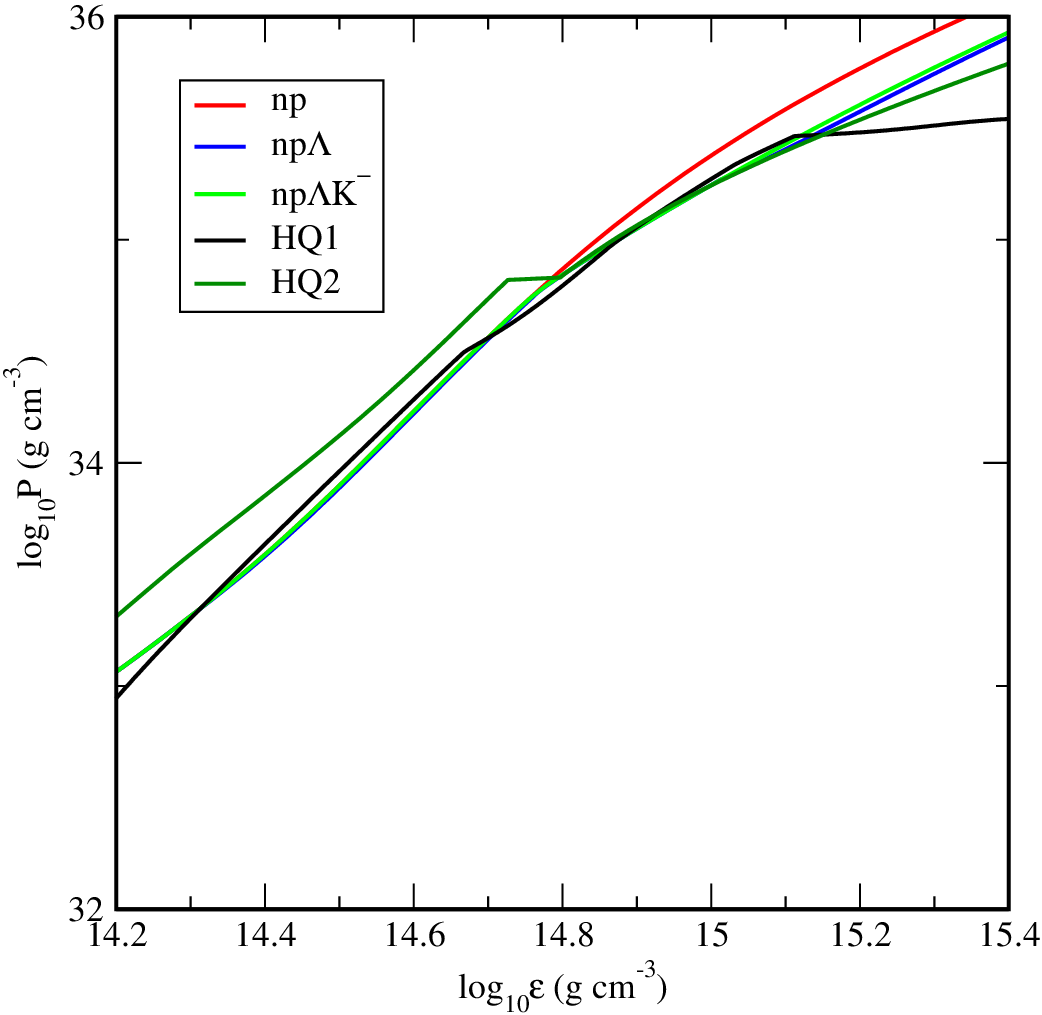}}
\caption{Pressure versus energy density is plotted for different compositions of
matter.}
\label{fig:1}       
\end{figure}
\end{center}
\begin{center}
\begin{figure}
\vspace*{1cm}       
\hspace{0.5cm}
\resizebox{0.4\textwidth}{!}{\includegraphics{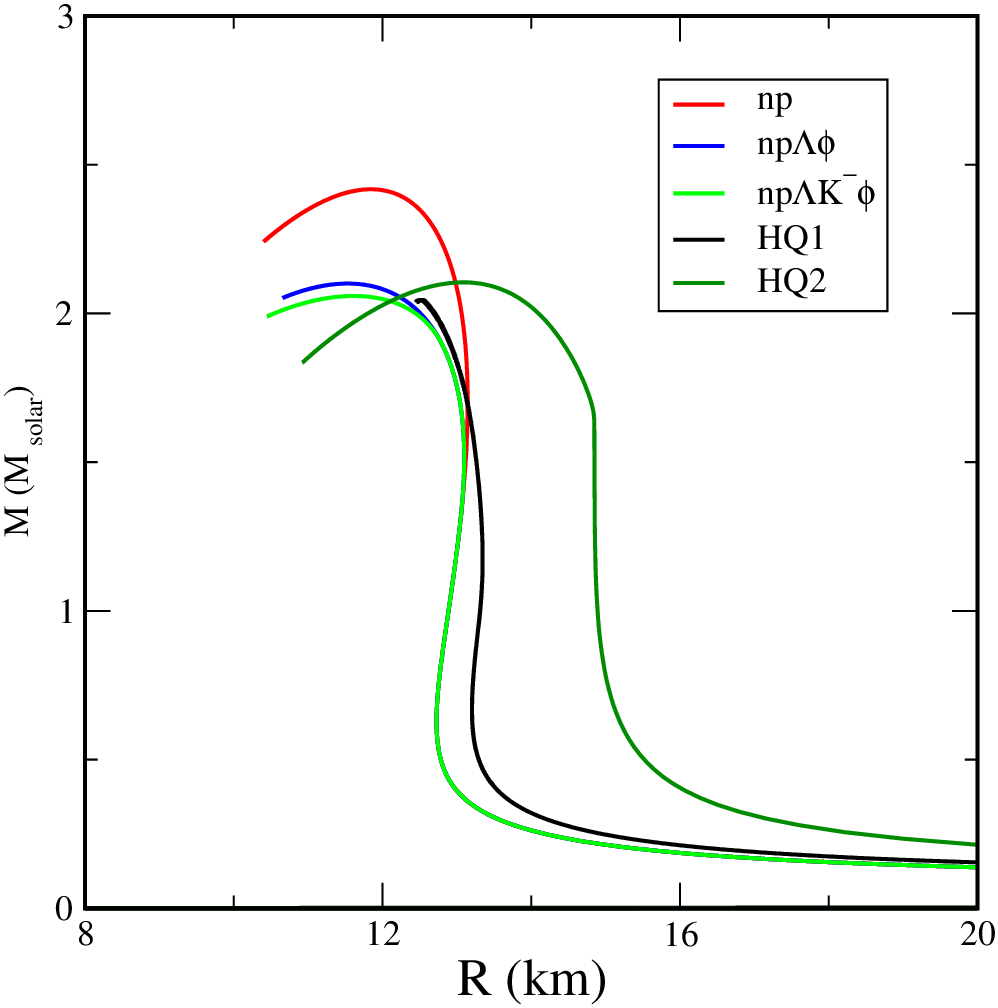}}
\caption{Mass versus radius is plotted for equations of state shown in Fig. 1.}
\label{fig:2}       
\end{figure}
\end{center}
The effective mass and in-medium energy of antikaons decrease with baryon 
density. The s-wave Bose-Einstein condensate of antikaons sets in when the
electron chemical potential is equal to the in-medium energy of antikaons. The 
phase transition from the hadronic to antikaon condensed phase could be either 
a first or second order phase transition. In the antikaon condensed phase, 
baryons are embedded in the condensate.
We treat the kaon-baryon interaction in the same footing as the baryon-baryon 
interaction in Eq. (\ref{had}). The Lagrangian density for (anti)kaons in the
minimal coupling scheme is \cite{char14,glen99,banik01},
\begin{equation}
{\cal L}_K = D^*_\mu{\bar K} D^\mu K - m_K^{* 2} {\bar K} K ~,
\end{equation}
where $K$ and $\bar K$ denote kaon and (anti)kaon doublets; the covariant 
derivative is
$D_\mu = \partial_\mu + ig_{\omega K}{\omega_\mu} 
+ ig_{\phi K}{\phi_\mu} 
+ i g_{\rho K}
{\mbox{\boldmath t}}_K \cdot {\mbox{\boldmath $\rho$}}_\mu$ and
the effective mass of antikaons is $m_K^* = m_K - g_{\sigma K} \sigma$.
Solving the equations of motion in the MFA, we obtain the EoS in the antikaon 
condensed phase. It is to be noted that kaon-meson couplings are not density
dependent. 
\begin{center}
\begin{figure}
\vspace*{1cm}       
\hspace{0.5cm}
\resizebox{0.4\textwidth}{!}{\includegraphics{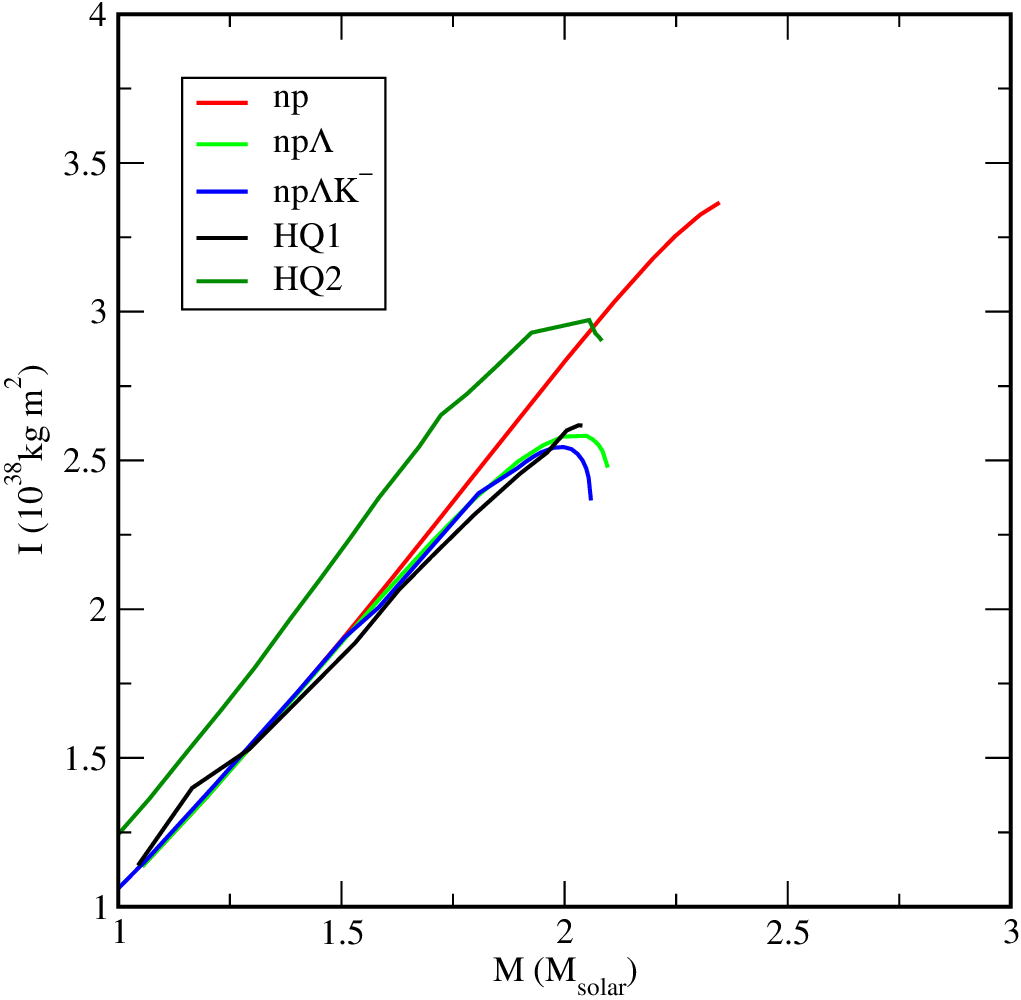}}
\caption{Moment of inertia is shown as a function of neutron star mass for 
different compositions of matter.}
\label{fig:3}       
\end{figure}
\end{center}
Next we consider two equations of state undergoing first order phase 
transition from hadronic matter to quark matter \cite{weber1,weber2,prc91}. 
In the first case, for the hadronic matter above 
saturation density, the DDRH model as described above is employed here. This 
hadronic phase includes all hyperons plus $\Delta$ resonance. A non-local 
extension of the Nambu-Jona-Lasinio model is adopted to describe the quark 
phase made of $u$, $d$ and $s$ quarks \cite{NJL}. The hadron-quark mixed phase
is governed by the Gibbs rules. This hadron-quark (HQ) EoS is denoted by 
HQ1 hereafter. In the other case, the hadronic phase is composed of only 
nucleons and described by the NL3 model whereas the effective bag model 
including quark interaction is adopted for the quark phase\cite{prc91}. The 
Maxwell construction is implemented for the mixed phase. We call this 
hadron-quark EoS as HQ2.   

\label{sec:1}
\subsection{Model of Rotating Neutron Stars}
Most of observed neutron stars rotate slowly. The fastest rotating neutron star
has a spin frequency of 716 Hz. Rotation or tidal field makes a neutron star
deformed. Here we study rigidly rotating neutron stars assuming 
a stationary, axisymmetric spacetime and the line element is given by 
\cite{BGSM,salgado}
\begin{equation}
ds^2 = - N^2 dt^2 + A^2(dr^2+r^2 d\theta^2) + B^2 r^2 \sin^2 \theta (d\phi - N^{\phi} dt)^2~,
\label{eq:metric}
\end{equation}
where metric functions $N,N^{\phi},A,B$ depend on coordinates $r$ and $\theta$.

The energy-momentum tensor for a perfect fluid which describes the matter, is  
\begin{eqnarray}
T^{\mu \nu} &=& (\varepsilon + P) u^{\mu} u^{\nu} + P g^{\mu \nu}~.
\label{emtensor}
\end{eqnarray}
Here the first term represents the matter contribution (perfect fluid), where 
$\varepsilon$ is the energy density, $P$ is the pressure and $u^{\mu}$ is the 
fluid four velocity. 

The equilibrium configurations for rotating neutron stars are determined by
solving the Einstein field equations. 
For rigid rotation, this is equivalent to solving a first integral of the 
equation of fluid stationary motion \cite{BGSM} 
\begin{equation}
H(r,\theta) + ln N - \ln \Gamma (r,\theta) = const~, 
\label{eq:first-integral} 
\end{equation}
where $\Gamma$ is the Lorentz factor of the fluid with respect to the Eulerian
observer and the fluid log-enthalpy is
\begin{equation}
 H = ln \left(\frac{\varepsilon + P}{n m_B}\right)~.
\end{equation}
Here $n$ and $m_B$ are baryon density and rest mass, respectively. \\

The first-integral (Eq. \ref{eq:first-integral}) is integrable within the 
3+1 formalism, where they reduce to Poisson-like partial differential equations.
These equations are then solved numerically using spectral scheme within the numerical library LORENE \cite{eric,lorene}. The input parameters
for the model are an Equation of State,the rotation frequency $\Omega$ and the 
central log-enthalpy $H_c$.
Global properties of rotating neutron stars such as 
gravitational mass, circumferential equatorial radius, angular momentum, moment of inertia and 
quadrupole moment are calculated within this formalism using the asymptotic behaviour of lapse function ($N$) and the component of the 
shift vector ($N^\phi$), in terms of metric potential and the source \cite{BGSM,salgado}. 
The gravitational mass, angular momentum and quadrupole moment are given respectively as \cite{salgado,Pappas,Friedman},
\begin{equation}
 M = \frac{1}{4 \pi} \int \sigma_{ln N} ~ r^2 \sin^2 \theta dr d\theta d\phi ~
\end{equation}
\begin{equation}
J = \int A^2 B^2 (E + p) U r^3 \sin^2 \theta dr d\theta d\phi ~.
\end{equation}
\begin{equation}
 Q = - M_2 - \frac{4}{3}\left(b+ \frac{1}{4}\right) M^3~, 
\end{equation}
  where,
\begin{equation}
M_2 = - \frac{3}{8\pi} \int \sigma_{ln N} \left(\cos^2 \theta - \frac{1}{3} \right) r^4 \sin^2 \theta dr d\theta d\phi ~
\end{equation}
Here, $\sigma_{ln N}$ is given by the RHS of Eq. $3.19$ of \cite{BGSM}, $U$ is the fluid four-velocity, $E = \Gamma^2(\varepsilon + p) - p$ and
$\Gamma = (1 - U^2)^{-1/2}$
and $b$ is defined by Eq. (3.37) of \cite{Friedman}.
Then, the moment of inertia of the rotating star is defined as,
\begin{equation}
 I :=  \left| \frac{J}{\Omega} \right|
\end{equation}
At lower rotation frequencies, this behaves almost as a linear relation.

\begin{figure}
\vspace*{1cm}       
\hspace{0.5cm}
\resizebox{0.4\textwidth}{!}{\includegraphics{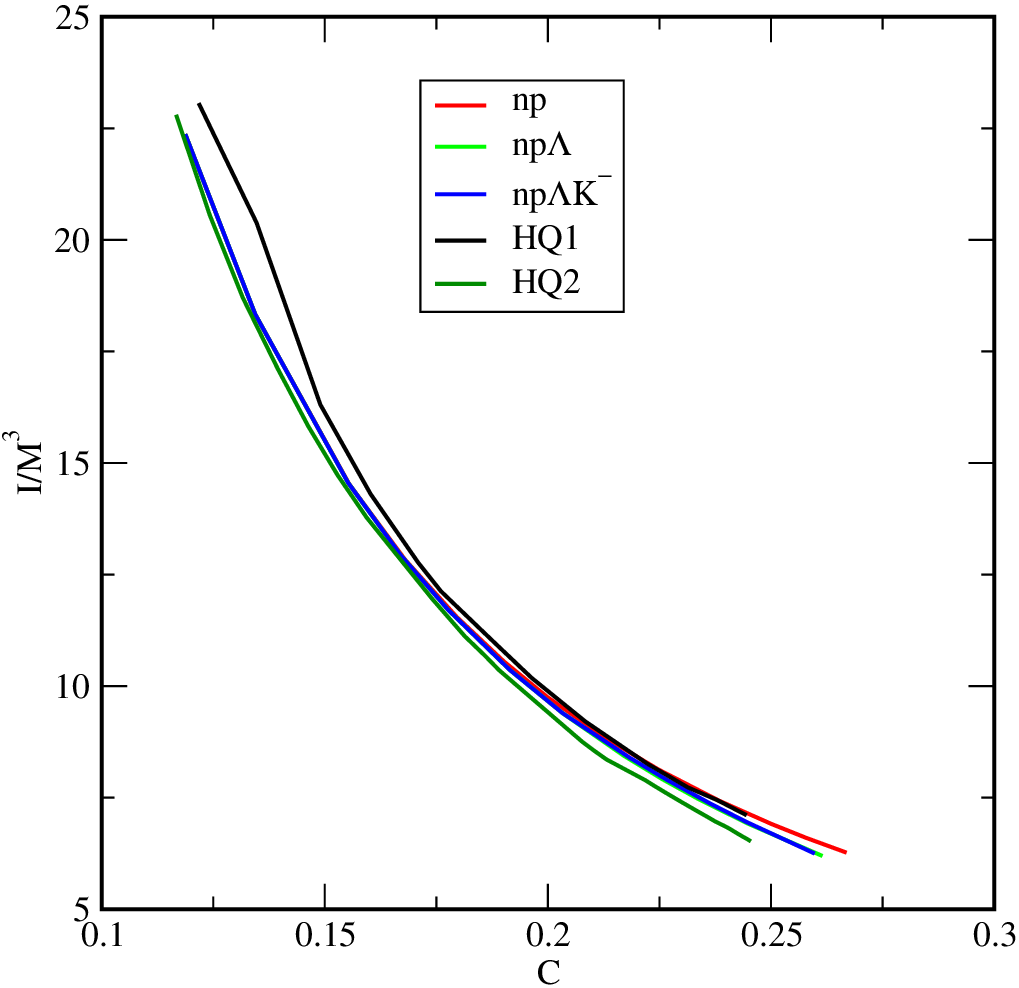}}
\caption{Dimensionless moment of inertia ($\bar I = I/M^3$) is plotted with 
compactness.}
\label{fig:4}       
\end{figure}
\subsection{Love Number, Moment of Inertia and Quadrupole Moment}

When a static, spherically symmetric star 
is under the influence of a static external quadrupolar tidal field
$E_{ij}$, it develops  a quadrupole moment $Q_{ij}$.
The tidal deformability $\lambda$ in the linear order is defined as
\begin{equation}
\lambda = - \frac{Q{ij}}{E_{ij}}
\end{equation}
The $\ell = 2$ dimensionless tidal Love Number ($k_2$) is related to the tidal deformability as 
\begin{equation}
k_2 = \frac{3}{2 R^5} \lambda
\end{equation}

We calculate the tidal love number following the prescription by Hinderer 
et al. \cite{hind,read}.
In the Regge-Wheeler gauge, a linear, static and even parity $\ell = 2, m = 0$ perturbation leads to the following form of deformed metric,
\begin{eqnarray}
ds^2 &=& - e^{2\Phi(r)} \left[1 + H(r)
Y_{20}(\theta,\varphi)\right]dt^2
\nonumber\\
& & + e^{2\Lambda(r)} \left[1 - H(r)
Y_{20}(\theta,\varphi)\right]dr^2
\nonumber \\
& & + r^2 \left[1-K(r) Y_{20}(\theta,\varphi)\right] \left( d\theta^2+ \sin^2\theta
d\varphi^2 \right),
\nonumber\\
& &
\end{eqnarray}
where,  $K'(r)=H'(r)+2 H(r) \Phi'(r)$. This leads to a second order differential equation for metric function $H$  
\begin{eqnarray}
H'' + H' \left(\frac{2}{r}+\Phi '-\Lambda'\right) + H \left(-\frac{6e^{2\Lambda}}{r^2}-2(\Phi ')^2\right.\nonumber\\
\left. +2\Phi'' +\frac{3}{r}\Lambda ' + \frac{7}{r}\Phi '- 2 \Phi'   \Lambda '+ \frac{f}{r}(\Phi '+\Lambda ')\right)=0.
\end{eqnarray}
where, $f= {d\epsilon}/{dp}$. We solve this equation by integrating outward from the center and using the asymptotic
behaviour of $H(r)$, we get the $\ell=2$ tidal love number,
\begin{eqnarray}
k_2 &=& \frac{8C^5}{5}(1-2C)^2[2+2C(y-1)-y]\nonumber\\
      & & \times\bigg\{2C[6-3y+3C(5y-8)]\nonumber\\
      & & +4C^3[13-11y+C(3y-2)+2C^2(1+y)]\nonumber\\
      & & +3(1-2C)^2[2-y+2C(y-1)] \ln(1-2C)\bigg\}^{-1}.
\end{eqnarray}
Here, we define $y=RH'(R)/H(R)$ and take compactness of the star, $C=M/R$.

It is to be noted that the tidal deformability parameter, moment of inertia and
quadrupole moment are dependent on the EoS individually. However, it was 
predicted that the relation among any two of them in the slow rotation
approximation was EoS independent \cite{yagi}. We use the following 
dimensionless quantities in this investigation. 
The dimensionless $\lambda$ is defined as 

\begin{equation}
\bar{\lambda} = \frac{\lambda}{M^5} 
\end{equation}

Similarly we introduce dimensionless moment of inertia and quadrupole moment as
$\bar I = \frac {I}{M^3}$ and $\bar Q = \frac{Q}{(M^3 (J/M^2)^2)}$. Here $Q$ is
compared with the Kerr solution quadrupole moment $J^2/M$. The dimensionless Q
is known as the Kerr factor. 
\label{sec:2}
\begin{figure}
\vspace*{1cm}       
\hspace{0.5cm}
\resizebox{0.4\textwidth}{!}{\includegraphics{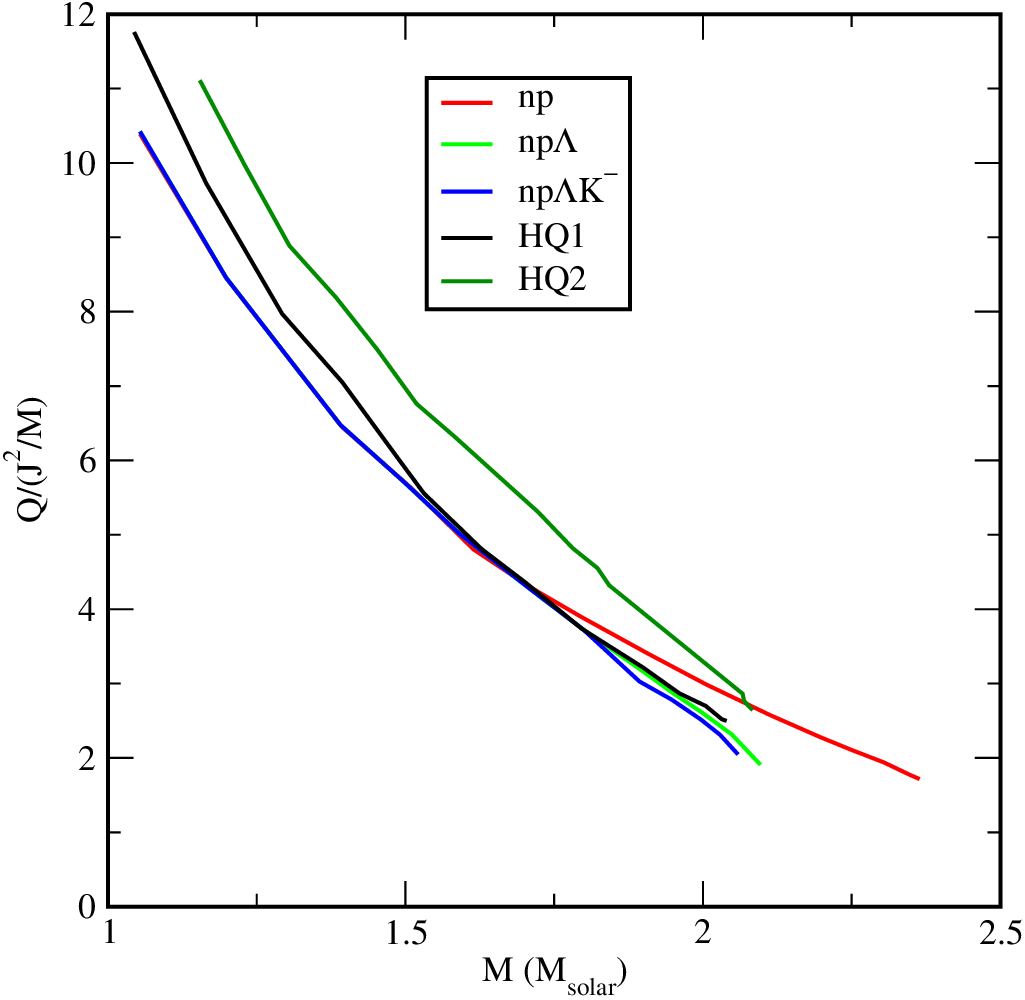}}
\caption{Dimensionless quadrupole moment with respect to the Kerr solution is 
plotted with neutron star mass.}
\label{fig:5}       
\end{figure}
\section{Results and Discussion}
We use the density dependent parameter set referred to as the 
DD2 set for nucleon-meson couplings \cite{banik14,typ10}. Hyperon-meson 
couplings are determined from the SU(6) symmetry relations and hypernuclear 
data \cite{sch96,weis}. Being lightest among all hyperons, $\Lambda$ hyperons 
might be populated first in dense matter. We consider only $\Lambda$ hyperons 
in this calculation. For $\Lambda$s, we consider a potential depth of -30 
MeV in normal nuclear matter and estimate the scalar meson coupling. 
Similarly, kaon-meson couplings are determined using the quark model, SU(3)
relations and kaon atomic data \cite{char14}. In this calculation, an attractive
potential depth of -120 MeV in normal nuclear matter is adopted for the 
determination of kaon-scalar meson coupling. Parameters of the quark phase are 
taken from Ref.\cite{weber1,prc91}.  With those parameter sets, we calculate 
various properties of rotating neutron stars as described in section 
\ref{sec:1} for four different compositions of matter namely neutron-proton 
matter (denoted as $np$), $\Lambda$ hyperon matter ($np\Lambda$), antikaon 
condensed matter including $\Lambda$ hyperons ($np\Lambda K^-$) and 
hadron-quark matter. 

Figure \ref{fig:1} exhibits the pressure ($P$) as a function of energy 
density ($\varepsilon$) known as the EoS for different compositions of matter.
It is evident from the figure that strange matter components such as $\Lambda$
hyperons, antikaons in the condensate or quarks make an EoS softer.Two 
kinks in both HQ EoSs show the beginning and end of the hadron-quark mixed phase
\cite{weber1,weber2,prc91}. Among the five EoSs, the HQ1 EoS involving 
hadron-quark phase transition is the softest EoS and $np$ 
matter has the stiffest EoS. We calculate static neutron star structures using 
those EoSs. The mass-radius relation is shown in Fig. \ref{fig:2}. It is 
found that maximum masses corresponding to $np$, $np\Lambda$, $np\Lambda K^-$, 
HQ1 and HQ2 EoS are 2.42, 2.1, 2.06, 2.04 and 2.1 M$_{solar}$, 
respectively. It demonstrates that all these EoSs are compatible with the 
2 M$_{solar}$ constraint.   

The chance for the measurement of moment of inertia has brightened with the 
detection of the relativistic pulsar binary PSR J0737-3039. In particular, the 
estimation of I for pulsar A in the double pulsar system might be possible in 
near future.  Masses of both pulsars in the double pulsar system had been 
accurately measured due to the knowledge of post Keplerian parameters. The 
precise measurement of I from observations along with the accurately known mass
of pulsar A might overcome the uncertainties in the determination of radius 
\cite{schu}. Here we present our results on the effects of exotic matter on 
moment of inertia. We are considering slowly rotating neutron stars with spin
frequency 100 Hz in this calculation using LORENE. The behaviour of 
moment of inertia with neutron star mass is
shown in Fig. {\ref{fig:3}}. It is noted that the moment of inertia 
corresponding to strange matter is lower than that of nuclear matter at higher 
neutron star masses. As we know the mass (1.337 M$_{solar}$) of pulsar A in 
PSR J0737-3039, we can predict the value of I from Fig. {\ref{fig:3}}.    

Dimensionless moment of inertia as defined in section \ref{sec:2} as a function
of compactness is shown in Fig. \ref{fig:4} for four EoSs. It is observed that
curves corresponding to $np$, $np\Lambda$ and $np\Lambda K^-$ merge before
the onset of $\Lambda$ hyperons or antikaon 
condensate. As soon as strange degrees of freedom appear in dense neutron star
matter, $\bar{I}$ for strange matter EoSs deviates from that of the 
nuclear matter case at higher compactness. However, it is noted that results
of $np\Lambda$ and $np\Lambda K^-$ cases show almost a universal behaviour. 
This may be attributed to the significant population of $\Lambda$ hyperons 
even after the appearance of antikaon condensate.  

We continue our investigation of quadrupole moment of rotating neutron stars 
having spin frequency of 100 Hz using LORENE. The behaviour of the 
quadrupole 
moment in comparison to the Kerr solution is shown with neutron star mass in 
Fig. \ref{fig:5}. It is found that the quadrupole moment decreases with mass 
and approaches the Kerr value around maximum neutron star 
mass. The stiffest EoS corresponding to the nuclear matter case is closest to
the Kerr solution. This was already noted by others \cite{yagi13,ruff,miller}. 
\begin{figure}
\vspace*{1cm}       
\hspace{0.5cm}
\resizebox{0.4\textwidth}{!}{\includegraphics{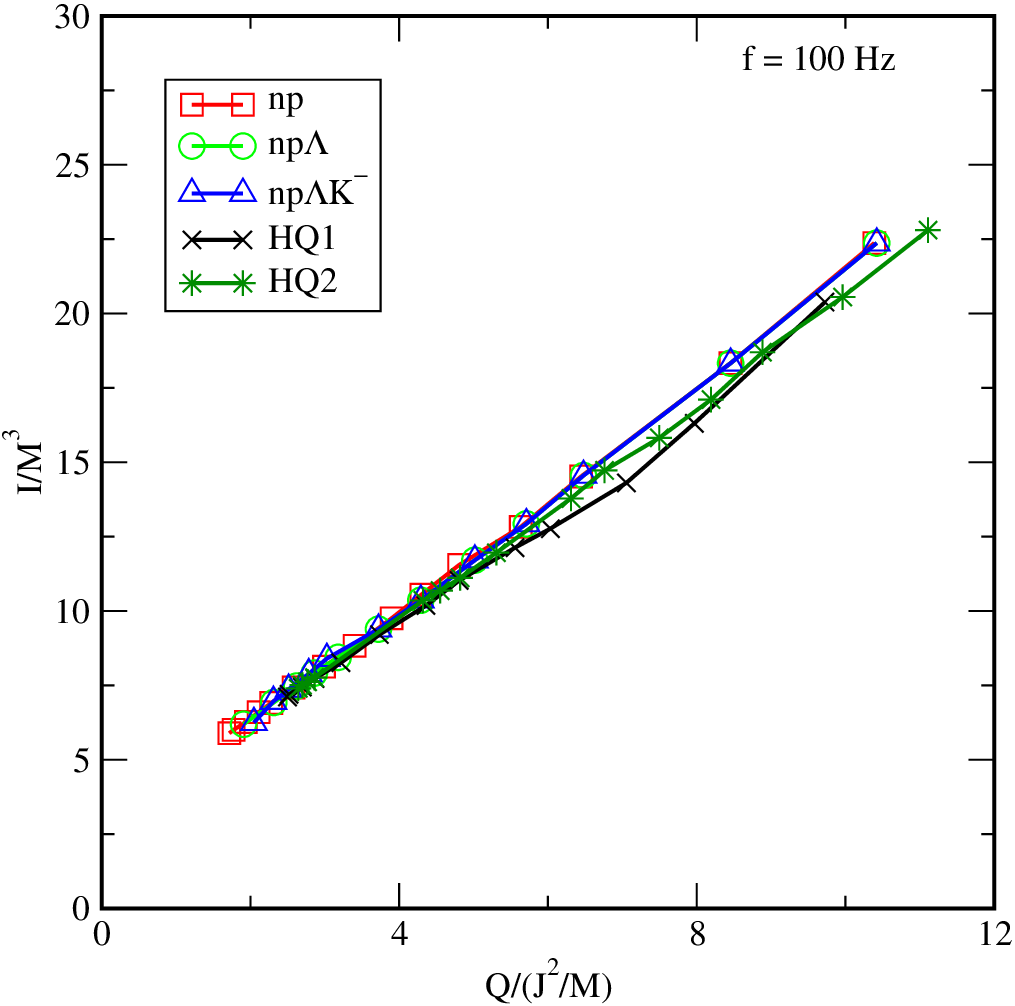}}
\caption{Dimensionless moment of inertia is shown as a function of dimensionless
quadrupole moment for rotational frequency 100 Hz.}
\label{fig:6}       
\end{figure}
\begin{figure}
\vspace*{1cm}       
\hspace{0.5cm}
\resizebox{0.4\textwidth}{!}{\includegraphics{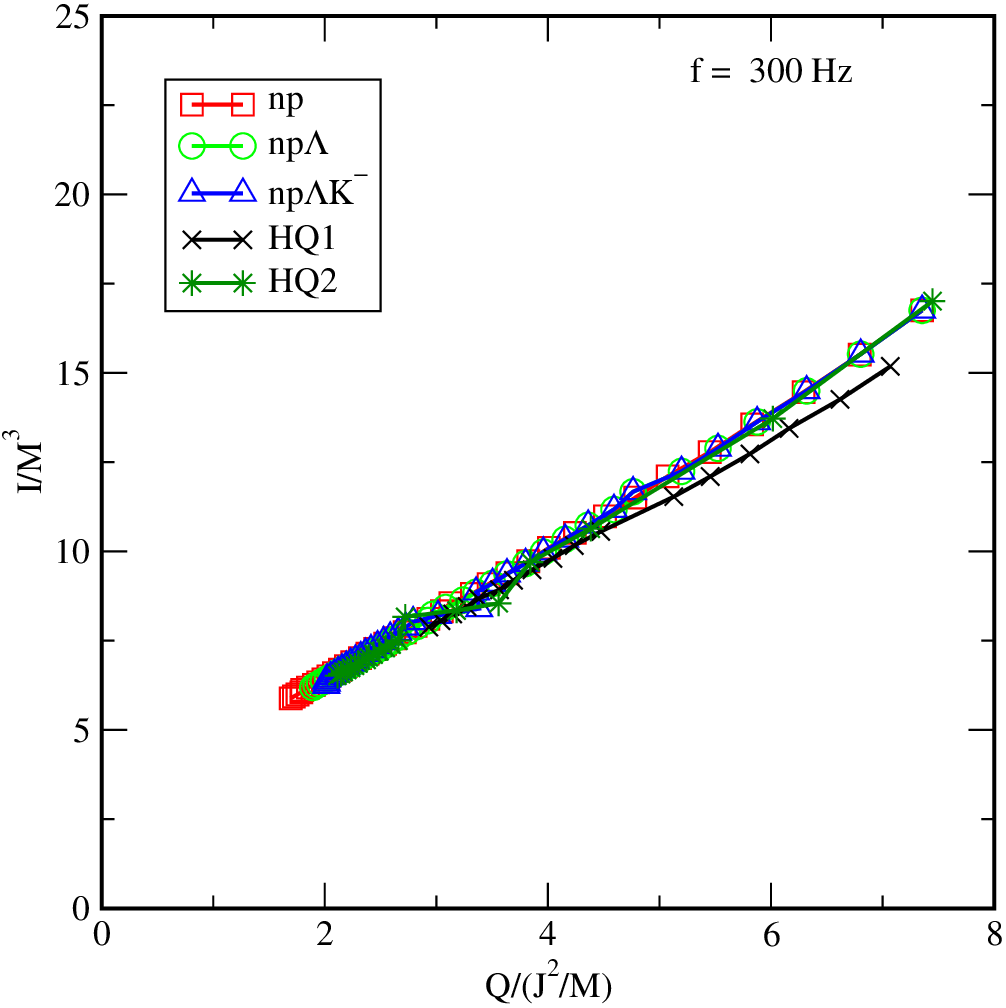}}
\caption{Same as Fig. 6 but for rotational frequency 300 Hz.} 
\label{fig:7}       
\end{figure}
We obtain the most interesting result when we plot dimensionless I versus 
dimensionless Q in Fig. \ref{fig:6}. All model calculated data 
corresponding to different EoSs fall on the same line except that of the HQ 
EoSs. Though I or Q is sensitive to EoS, their relationship exhibit universality
for $np$, $np\Lambda$ and $np\Lambda K^-$ EoS. This kind of universal 
relation was first predicted by Yagi and Yunes \cite{yagi}. This holds good 
for other observable quantities also. When the ratio of critical mass i.e., 
the maximum mass at the mass-shedding limit and maximum mass for static
neutron stars are plotted with normalised angular momentum with respect to 
maximum angular momentum, it also shows a universal relation \cite{breu,ssl}. 
However, this universality appears to be lost when we look at both HQ 
EoSs. This may be attributed to a first order hadron-quark phase 
transition in HQ1 and HQ2.

\begin{figure}
\vspace*{1cm}       
\hspace{0.5cm}
\resizebox{0.4\textwidth}{!}{\includegraphics{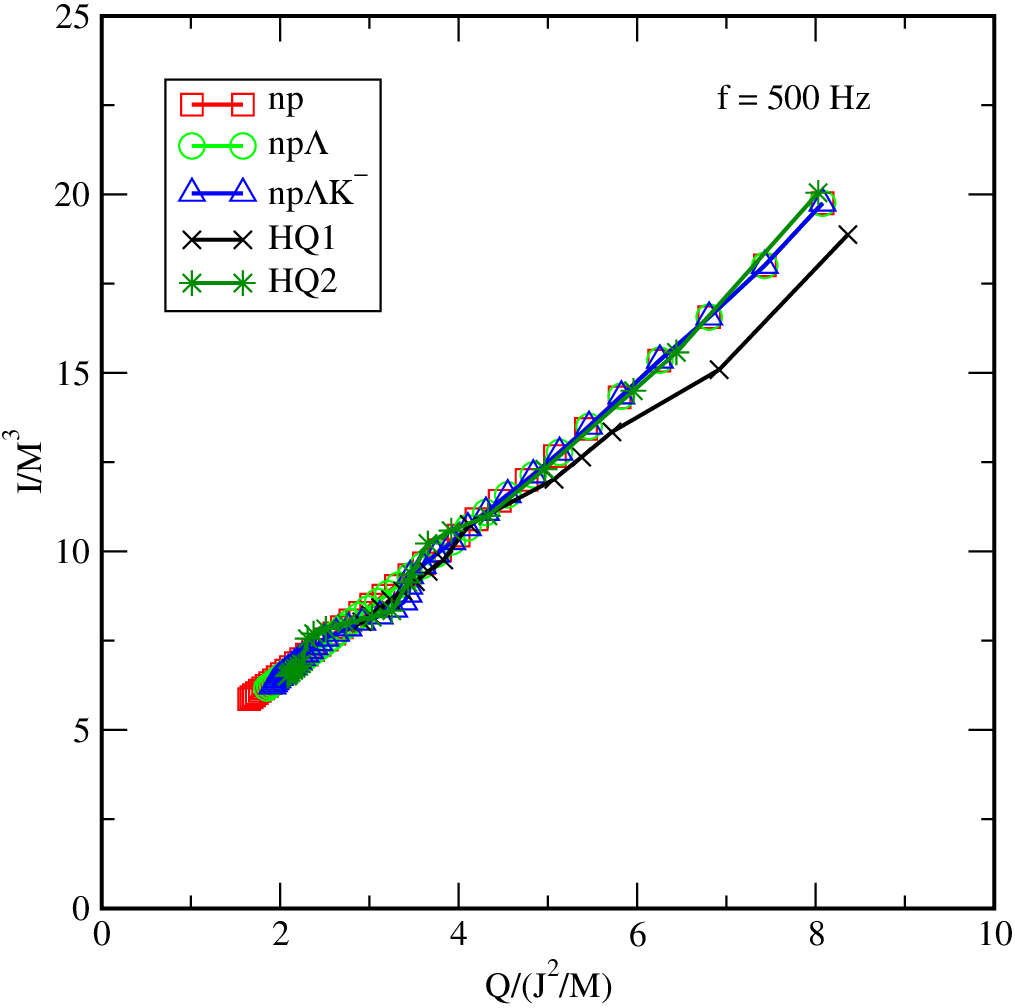}}
\caption{Same as Fig. 6 but for rotational frequency 500 Hz.} 
\label{fig:8}       
\end{figure}

We further investigate the role of first order phase transition 
on the I-Q universality relation at higher rotational frequencies. It is a 
well known fact that the rotation drives compositional changes in neutron 
stars \cite{weber1}. This has significant implications for
the hadron-quark phase transition in rotating neutron stars. As a neutron star
spins down, its central density increases from below. When the density exceeds
the threshold value, a hadron-quark phase transition sets in. We study I-Q 
relation for 300 and 500 Hz to understand the above effect and the results
are demonstrated in Fig. \ref{fig:7} and Fig. \ref{fig:8}. For HQ2 EoS,
the hadron-quark phase transition disappears at those frequencies and the 
universality is retained. However, the imprint of the phase transition
can be found even at higher rotational frequencies for HQ1 case. This may be 
attributed to the fact that HQ1 EoS has a much wider mixed phase compared with
that of HQ2 EoS. 

\begin{figure}
\vspace*{1cm}       
\hspace{0.5cm}
\resizebox{0.4\textwidth}{!}{\includegraphics{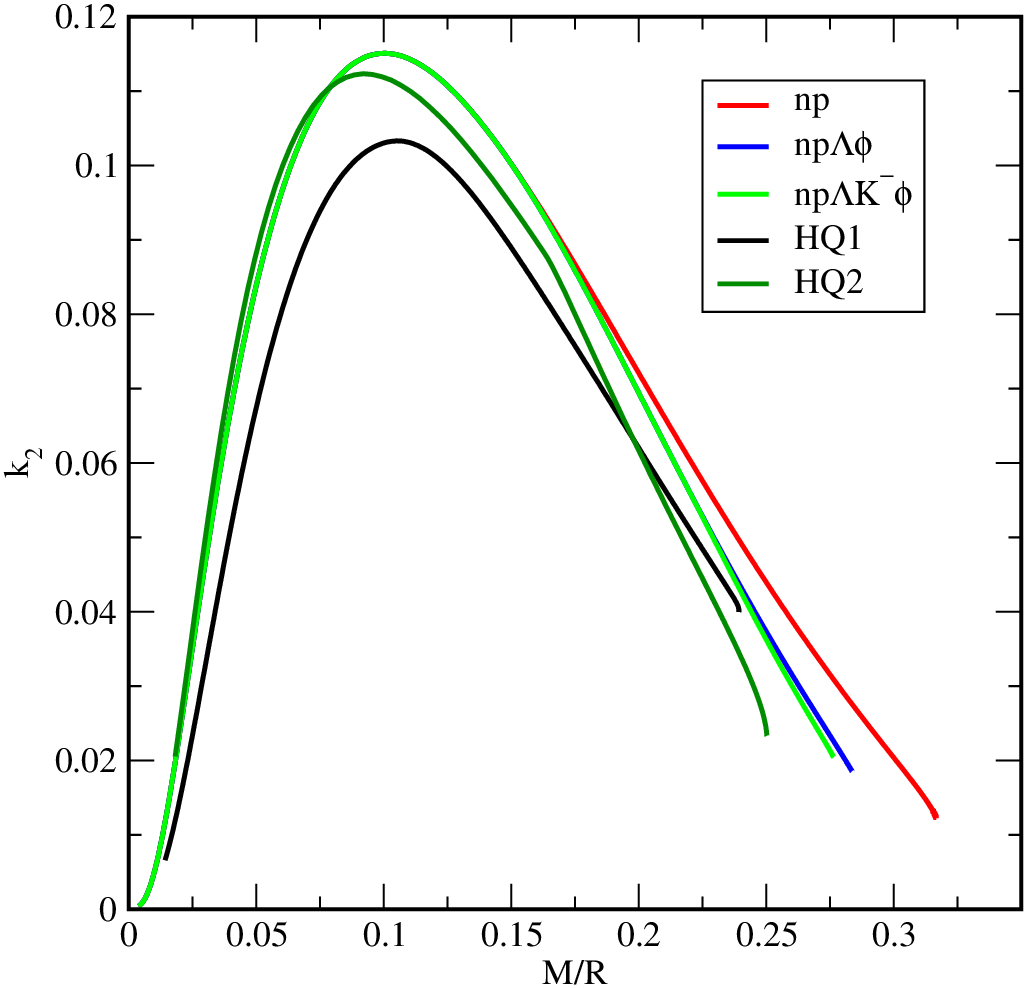}}
\caption{Love number is plotted with compactness of neutron star.}
\label{fig:9}       
\end{figure}

\begin{figure}
\vspace*{1cm}       
\hspace{0.5cm}
\resizebox{0.4\textwidth}{!}{\includegraphics{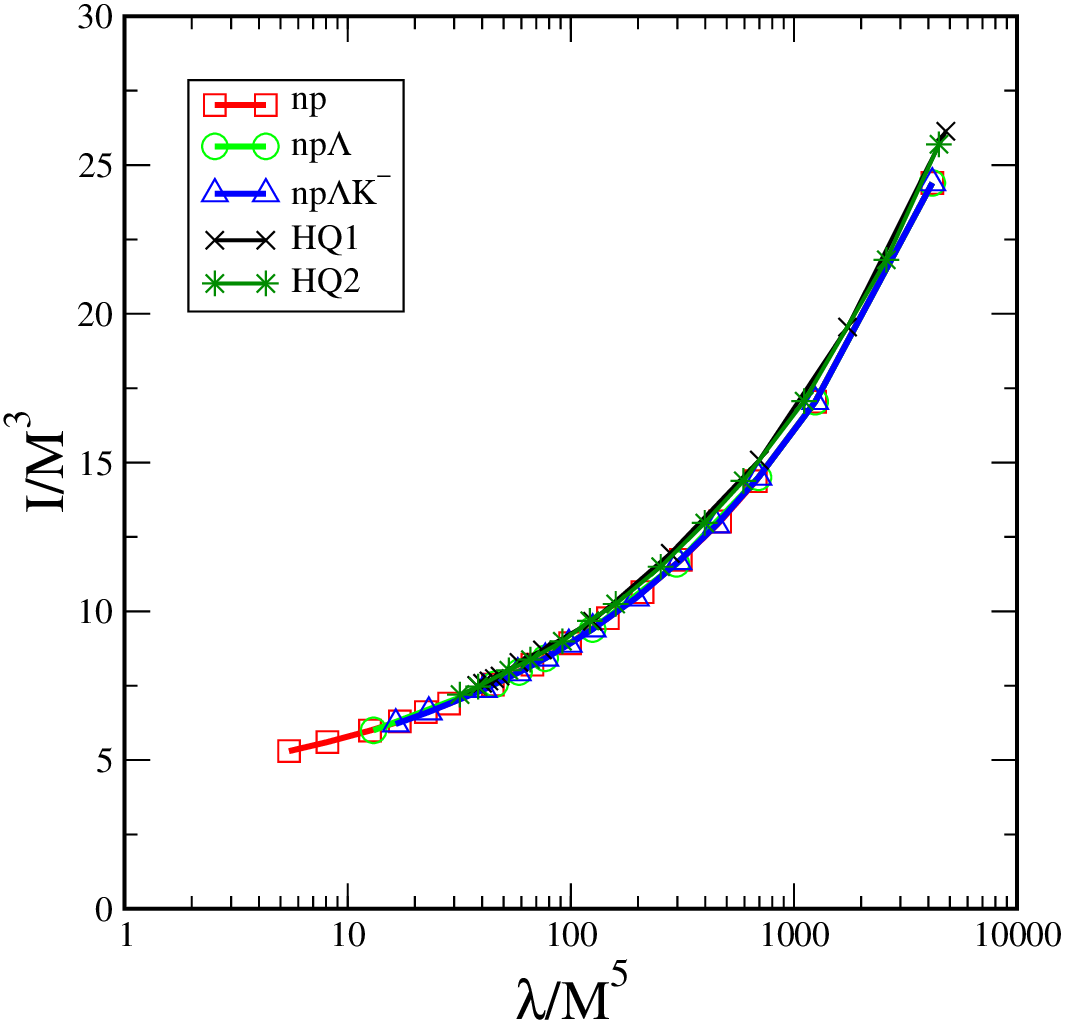}}
\caption{Dimensionless moment of inertia is shown as a function of dimensionless
deformability parameter ($\bar{\lambda}$)}
\label{fig:10}       
\end{figure}

Next we study the relation between I and Love number for strange matter EoSs 
adopting slow rotation approximation of Hartle and Thorne \cite{yagi13}.
First we plot $\ell = 2$ love number ($k_2$) with compactness ($C = M/R$) in 
Fig.\ref{fig:9} for $np$, $np\Lambda$, $np\Lambda K^-$ and HQ matter. The love
number measures how easy or difficult to deform a neutron star. For each 
EoS, love number decreases as the compactness increases. This implies  
that more the star is compact, less is its love number. Furthermore, this 
effect is more pronounced in softer EoSs i.e. $np\Lambda$, $np\Lambda K^-$ and 
HQ than that of $np$ matter. The love number approaches zero for black holes 
for which compactness is 0.5. 
Figure \ref{fig:10} describes the behaviour of dimensionless moment of inertia 
with dimensionless tidal deformability parameter $\bar{\lambda}$ as defined in 
Sec. \ref{sec:2} for $np$, $np\Lambda$, $np\Lambda K^-$ and HQ EoS. Like Fig. 
\ref{fig:6}, we find again a universal relation in $\bar{I}$-$\bar{\lambda}$ 
for three hadronic EoSs. However, the data of HQ EoSs are deviating from 
the universal relation.  
It is now evident that Fig. \ref{fig:6} and Fig. \ref{fig:10} together produce 
I-Love-Q relations which are not sensitive to the compositions and EoS of 
neutron star matter without a first order phase transition. In those
cases, neutron star matter below the saturation density is well 
constrained by nuclear physics experiments in laboratories. All EoSs in this 
density regime should behave in the same fashion and lead to universal 
relations. However, this is not true at higher densities where many new degrees
of freedom appear in the form of hyperons, antikaon condensate and quarks. We
find significant deviations in EoSs in that density regime. It is then a puzzle
to understand what drives the universality when the neutron star compactness
increases. This has been attributed to the isodensity contours in neutron
stars which are approximately elliptically self similar \cite{yagi17,val}. It 
is worth mentioning here this kind of universal relations among
I and Q were also found in rotating protoneutron stars \cite{val,hem}.          

\label{sec:3}
\section{Conclusions}
We study properties of rotating neutron stars using LORENE as well as 
the Hartle-Thorne prescription of slowly rotating neutron stars. We exploit
different EoSs involving hyperons, Bose-Einstein condensate of antikaons and
quarks. 
Next we investigate the behaviour of moment of inertia, quadrupole moment and 
their dimensionless counter parts with gravitational mass and compactness of 
neutron stars. Furthermore, we study the Love number for different EoSs.    
We find that all those quantities are dependent on EoSs. It is also noted that
the tidal Love number approaches a smaller value for large compactness. 
Similarly, the quadrupole moment of a rotating neutron star moves closer towards
the Kerr value of a black hole for maximum mass neutron stars. 

We further investigate the relations among dimensionless moment of inertia,
quadrupole moment and tidal deformability parameter. The universal I-Q and
I-Love number relations are observed in our calculation for EoS including 
hyperons and antikaon condensate as predicted by
Yagi and Yunes \cite{yagi}. However, we have shown for the first time that 
the universality in I-Q relation is violated for HQ EoSs undergoing a first 
order phase transition.

\vspace{1cm}

\leftline {\bf Acknowledgements}
DB has fond memories of his interactions with Walter Greiner.

\end{document}